\documentstyle[amssymb,preprint,aps]{revtex}

\begin{document}
\title{Angular momentum spatial distribution symmetry breaking in Rb by an external
magnetic field}
\author{Janis Alnis, Marcis Auzinsh\thanks{%
Corresponding author, Fax +371-7820113, e-mail mauzins@latnet.lv}}
\address{Department of Physics, University of Latvia, 19 Rainis blvd., Riga LV-1586,\\
Latvia}
\date{\today}
\maketitle

\begin{abstract}
Excited state angular momentum alignment -- orientation conversion for atoms
with hyperfine structure in presence of an external magnetic field is
investigated. Transversal orientation in these conditions is reported for
the first time. This phenomenon occurs under Paschen Back conditions at
intermediate magnetic field strength. Weak radiation from a linearly
polarized diode laser is used to excite Rb atoms in a cell. The laser beam
is polarized at an angle of $\pi /4$ with respect to the external magnetic
field direction. \ Ground state hyperfine levels of the $5S_{1/2}$ state are
resolved using laser-induced fluorescence spectroscopy under conditions for
which all excited $5P_{3/2}$ state hyperfine components are excited
simultaneously. Circularly polarized fluorescence is observed to be emitted
in the direction perpendicular to both to the direction of the magnetic
field ${\bf B}$ and direction of the light polarization ${\bf E}$. The
obtained circularity is shown to be in quantitative agreement with
theoretical predictions.
\end{abstract}

\section{Introduction}

In the absence of external forces an ensemble of unpolarized atoms can only
be aligned \cite{alignment} by linearly polarized light. The fact that the
atoms are only aligned implies that, although the magnetic sublevels of
different $\left| m_{J}\right| $ are populated unequally, magnetic sublevels
of $+m_{J}$ and $-m_{J}$ are equally populated. For this reason, atoms
excited by linearly polarized laser radiation are not expected to produce
circularly polarized fluorescence. In the presence of external forces,
however, excitation by linear polarized light can produce an orientated
population\ of atoms (with different $+m_{J}$ and $-m_{J}$ populations.)
This effect, called alignment -- orientation conversion, was predicted and
experimentally observed in the late sixties in the anisotropic collisions of
initially aligned atoms \cite{lomb1,reb1,lomb2,man1}. Later an electric
field was also shown to induce alignment -- orientation conversion \cite
{lomb3}.\ Electric-field-induced alignment -- orientation conversion has
since been studied in great detail\cite{auz1}.

Contrary to the case of an electric field, linear perturbation by a magnetic
field is not able to orient an initially aligned angular momentum
distribution. This inability to induce alignment -- orientation conversion
is a result of the reflection symmetry of axial vector fields. This symmetry
can be broken if, in addition to the linear Zeeman effect, there exists
nonlinear dependencies of the magnetic sublevel energies on the field
intensity and the magnetic quantum numbers $m_{J}$. Such nonlinear
perturbations can have a variety of causes including predissociation\cite
{vig1,vig2,auz2,auz3} and hyperfine interaction. Alignment -- orientation
conversion as a result of hyperfine interaction in a magnetic field in
context of nuclear spin $I=1/2$ was studied by J. Lehmann for the case of
optically pumped cadmium in a magnetic field\cite{leh1,leh2}. W. Baylis
described the same effect in sodium \cite{bey1}. The first experiment to
detect directly a net circular polarization of fluorescence from an
initially aligned excited state in an external magnetic field was reported
by M. Krainska -- Miszczak \cite{kra1}. \ In this work the optical pumping
of $^{85}$Rb by a $\pi $-polarized D$_{2}$ line was studied. This effect was
also examined by X. Han and G. Schinn in sodium atoms\cite{han1}. They
describe this alignment - orientation conversion process as resulting from
hyperfine-$F$-level mixing in an external magnetic field and the
interference of different excitation -- decay pathways in such mixed levels.

In all above cases, a joint action of the magnetic field and hyperfine
interaction creates different population of magnetic sublevels $+m_{F}$ and $%
-m_{F}$ of hyperfine levels $F$. This means that {\em longitudinal}
orientation of atoms along the direction of an external magnetic field is
created. Recently it was predicted that joint action of a magnetic field and
hyperfine interaction from an initially aligned ensemble would create {\em %
transverse} orientation of angular momentum\ of atoms or molecules \cite
{auz4}. Transverse orientation implies orientation in a direction
perpendicular to the external magnetic field ${\bf B}$. In this particular
case magnetic sublevels $+m_{F}$ and $-m_{F}$ are equally populated, but
orientation is a result of coherence between pairs of wave functions of
magnetic sublevels $m_{F}$ with $\Delta m_{F}=1$. \ Creation of transverse
orientation is achieved if the excitation light polarization vector is
neither parallel nor perpendicular to the external magnetic field direction
with the largest effect occurring for the case of a light polarization --
magnetic field angle of $\pi /4$. \ In a previous paper\cite{auz4},
parameters of the NaK molecule were used for numerical simulations of
orientation and fluorescence circularity signals. We found that transverse
orientation only occurred when the rotational angular momentum $J$ is small
enough to be comparable with the nuclear spin $I$. For levels with larger
angular momentum quantum number, the magnitude of created orientation was
found to decrease rapidly.

Previously transverse alignment -- orientation conversion was studied in
detail for the case of an external electric field\cite{auz1}. \ In this case
the conversion occurs with or without hyperfine interaction. \ In this paper
we report the first experimental observation to our knowledge of alignment
-- orientation conversion that creates net transverse orientation of atoms
with hyperfine structure in an external magnetic field. \ As we will show,
this effect is interesting not only as a new way to create orientated atoms,
but also can be used to increase the accuracy with which constants related
to the hyperfine interaction can be determined.

\section{Theoretical description}

The general scheme how transverse orientation of angular momentum is created
from an aligned ensemble of atoms is the following: \ Initial alignment, for
example by absorption of a linearly polarized light, is created at some
non-zero acute angle with respect to the direction of an external-field (in
this case a ${\bf B}$-field.) \ The optimum angle is $\pi /4$, but the
effect will take place at any angle that differs form $0$ and $\pi /2$. \
The perturbing field together with the hyperfine interaction causes
unequally spaced magnetic sublevel splittings. \ Under these conditions,
angular momenta orientation at the direction perpendicular to the direction
of the external field is created \cite{auz7}. \ A semiclassical
interpretation of this effect in terms of angular momentum precession in an
external field can be found in a previous publication \cite{auz1}. \ In this
vectorial model, alignment -- orientation conversion is the result of a
different precession rate for different orientations of angular momentum
with respect to the external field. \ In what follows we explain this
transverse-orientation in terms of a accurate quantum mechanical model.

In the present study we exploit laser excitation of pure isotopes of Rb
atoms from their ground state $5S_{1/2}$ to the first excited state $%
5P_{3/2} $ (resonance D$_{2}$ line) (see inset Figure \ref{Rb_85_lin}.) \ 
\label{se_th}The two most common naturally occurring isotopes of Rubidium
are $^{85}$Rb(72.15 \%, nuclear spin $I=5/2$) and $^{87}$Rb(27.85\%, nuclear
spin $I=3/2$.) \ As a result of hyperfine interactions, the ground-state
level of $^{85}$Rb is split into components with total angular momentum
quantum numbers $F_{i}=2$ and $F_{i}=3$ and the ground-state level of $^{87}$%
Rb is split into components with total angular momentum quantum numbers $%
F_{i}=1$ and $2$. The ground-state-level splittings for $^{85}$Rb and $^{87}$%
Rb are approximately $3$ GHz and $6$ GHz respectively. In contrast, the four
excited state hyperfine components are separated by only several hundred MHz
(see Figures \ref{Rb_85_levels} and \ref{Rb_87_levels}).

Excited-state-hyperfine structure in absorption is not resolved due to
Doppler broadening and laser-frequency jittering. To make an accurate signal
modeling assuming broad line excitation, laser frequency is modulated by a
few hundred MHz superimposing a $10$ kHz sine wave on laser current. This
allows accurate modeling to be done assuming that the excitation radiation
is broad enough to excite all hyperfine components of the excited state
without frequency selection, yet narrow enough to completely resolve the two
ground-state components.

Magnetic field caused mixing takes place between sublevels with different
total angular momentum $F_{e}$, but with identical magnetic quantum numbers $%
m_{F}$. \ Only levels of identical $m_{F}$ mix because, as far as the
magnetic field possesses axial symmetry, the magnetic quantum number $m_{F}$
remains a good quantum number. \ However, levels of different $F_{e}$ mix
because an intermediate strength magnetic field partially decouples the
electronic angular momentum $J_{e}$ and nuclear spin $I$. As a consequence $%
F_{e}$ ceases to be a good quantum number. \ The fact the $m_{F}$ remains a
good quantum number whereas $F_{e}\,$does not is important to the
interpretation of the data.

A convenient way to describe excited state atoms is by means of a quantum
density matrix $^{kl}f_{mm^{\prime }}$\cite{auz5}. \ Upper indices
characterize atomic states in a magnetic field. In the weak field limit
these states correspond to hyperfine levels $F_{e}$. Lower indices
characterize magnetic quantum numbers. \ We consider an atom possessing the
hyperfine structure which is placed in an external magnetic field. We
further assume that this atom absorbs laser light polarized in the direction
characterized by light electric field vector ${\bf E}_{exc}$. In this
situation the density matrix that characterizes coherence between magnetic
sublevels with quantum numbers $m$ and $m^{\prime }$ is given as\cite{auz6}

\begin{equation}
^{kl}f_{mm^{\prime }}=\frac{\widetilde{\Gamma }_{p}}{\Gamma +i^{kl}\Delta
\omega _{mm^{\prime }}}\sum_{j\mu }\left\langle \gamma _{k}m\right| \widehat{%
{\bf E}}_{exc}^{\ast }{\bf \cdot }\widehat{{\bf D}}\left| \eta _{j}\mu
\right\rangle \left\langle \gamma _{l}m^{\prime }\right| \widehat{{\bf E}}%
_{exc}^{\ast }{\bf \cdot }\widehat{{\bf D}}\left| \eta _{j}\mu \right\rangle
^{\ast }.  \label{eq1}
\end{equation}
Here $\widetilde{\Gamma }_{p}$ is a reduced absorption rate, $\Gamma $ is
the excited state relaxation rate and $^{kl}\Delta \omega _{mm^{\prime
}}=(^{\gamma _{k}}E_{m}-^{\gamma _{l}}E_{m^{\prime }})/\hbar $ is the energy
splitting of magnetic sublevels $m$ and $m^{\prime }$ belonging to the
excited state levels $k$ and $l$. Magnetic quantum numbers of the ground
state level $\eta _{j}$ are denoted by $\mu $ and magnetic quantum numbers
of the excited state level $\gamma _{k}$ by $m$ and $m^{\prime }$.

In an external magnetic field, ground- and excited- state levels $\eta _{j}$
and $\gamma _{k}$ are not characterized by a total angular momentum quantum
numbers $F_{i}$ and $F_{e}$, but are instead mixtures of these states: 
\begin{equation}
\left| \gamma _{k}m\right\rangle
=\sum_{F_{e}=J_{e}-I}^{F_{e}=J_{e}+I}C_{kF_{e}}^{\left( e\right) }\left|
F_{e},m\right\rangle ,\qquad \left| \eta _{j}\mu \right\rangle
=\sum_{F_{i}=J_{i}-I}^{F_{i}=J_{i}+I}C_{jF_{i}}^{\left( i\right) }\left|
F_{i},\mu \right\rangle .  \label{eq2}
\end{equation}
The wave-function-expansion coefficients $C_{kF_{e}}^{\left( e\right)
},C_{jF_{i}}^{\left( i\right) }$ \ represent the mixing of field free
hyperfine state wave functions by the magnetic field. These expansion
coefficients along with the magnetic sublevel energy splittings $^{kl}\Delta
\omega _{mm^{\prime }}$ can be obtained by a standard procedure of
diagonalization of a Hamilton matrix that contains both the diagonal
hyperfine elements and the off-diagonal magnetic field interaction elements
(see for example \cite{auz4}.)

There are several methods how to tell whether or not a particular atomic
state described by a density matrix (\ref{eq1}) possesses orientation. One
possibility is to expand this matrix over the irreducible tensorial
operators. Then those expansion coefficients can directly be attributed to
the alignment and orientation of the atomic ensemble \cite{auz5,var1,sob1}.
Alternatively, one may calculate directly the fluorescence circularity rate
in spontaneous transitions from a particular excited state of an atom: 
\begin{equation}
C=\frac{I({\bf E}_{right})-I({\bf E}_{left})}{I({\bf E}_{right})+I({\bf E}%
_{left})}  \label{eq3}
\end{equation}
Observed circularity of the fluorescence in a specific direction can differ
from zero only for the case that the ensemble of atoms possesses overall
orientation in this direction \cite{auz5}.\ $\ I({\bf E}_{right})$ and $I(%
{\bf E}_{left})$ are intensities of two fluorescence components with
opposite circularity. \ We choose to calculate this expected circularity
rate because it is the experimental measure used to register the appearance
of orientation in an ensemble of atoms (see for example \cite{auz7}).

We consider the case that spontaneous emission is detected without\
hyperfine-state resolution. \ The intensity of the fluorescence with
definite polarization characterized by a vector ${\bf E}_{f}$ in a
spontaneous transition from an excited state $J_{e}$ characterized by a set $%
\gamma _{k}$ of levels in an external field to the ground state $J_{f}$
characterized by a set $\eta _{j}$ of levels can be calculated according to
a previous work\cite{auz6} as 
\begin{equation}
I\left( {\bf E}_{f}\right) =I_{0}\sum_{mm^{\prime }\mu
}\sum_{klj}\left\langle \gamma _{k}m\right| \widehat{{\bf E}}_{f}^{\ast }%
{\bf \cdot }\widehat{{\bf D}}\left| \eta _{j}\mu \right\rangle \left\langle
\gamma _{l}m^{\prime }\right| \widehat{{\bf E}}_{f}^{\ast }{\bf \cdot }%
\widehat{{\bf D}}\left| \eta _{j}\mu \right\rangle ^{\ast kl}f_{mm^{\prime
}}.  \label{eq4}
\end{equation}
To find the circularity rate $C,$ one needs to not only determine the matrix
elements appearing in (\ref{eq1}) and (\ref{eq4}), but also the hyperfine
level splitting and magnetic sublevel mixing coefficients. In Figure \ref
{Rb_85_levels} the hyperfine energy level splitting of the first excited
state $5P_{3/2}$ for $^{85}$Rb is presented. In these calculations the
following published\cite{ari1} hyperfine splitting constants and magnetic
moment for the rubidium atom in its first excited state are used: $a=25.009$
MHz, $b=25.83$ MHz, $g_{J}=-1.3362$, $g_{I}=0.000293$.

\ In Figure \ref{Rb_85_levels} level crossing positions for magnetic
sublevels with $\Delta m_{F_{e}}=2$ are indicated by circles and crossings
with $\Delta m_{F_{e}}=1$ by squares. \ At values of magnetic field strength
for which coherently excited magnetic sublevels undergo a level crossing, $%
^{kl}\Delta \omega _{mm^{\prime }}=0$ the prefactor appearing in Equation (%
\ref{eq1})\ becomes large. \ This leads to resonance behavior of\ the
observed signal. \ For case of excitation with linearly polarized light, the
intensity of the resonance depends upon the angle between polarization
direction of the laser light and external magnetic field direction. If the
angle between these directions is $0$, different magnetic sublevels are
differently populated but no coherence is created in the ensemble. If the
angle is $\pi /2$ coherence is created between magnetic sublevels with $%
\Delta m_{F_{e}}=2$. \ If the angle differs form $0$ and $\pi /2$ then
magnetic sublevels with $\Delta m_{F_{e}}=1$ and $2$ \cite{auz7} are excited
coherently. \ This $\Delta m_{F_{e}}=1$ coherence is required for transverse
orientation.

We now consider the fluorescence circularity enhancement due to $\Delta
m_{F_{e}}=1$ level crossing for the case\ that the linear polarization and
external field meet at an angle of $\pi /4$ (inset Figure \ref{Rb_85_circ}.)
The circularity $C$ is calculated assuming an excited state relaxation rate 
\cite{sva1} $\Gamma =3.8\times 10^{7}$ s$^{-1}$ and observation along an
axis normal to the plain containing the external field ${\bf B}$ and the
polarization vector ${\bf E}_{exc}$. The smooth lines of Figure \ref
{Rb_85_circ} give the expected signals for both resolved absorption lines.
Both signals are maximum at an approximate magnetic field strength of $10$
G. \ For both absorption lines we calculate a total fluorescence circularity
with unresolved hyperfine components in a transition back to the ground
state $5S_{1/2}$. The resonance peak is more pronounced for the $%
F_{i}=2\longrightarrow F_{e}$ absorption transition than for the $%
F_{i}=3\longrightarrow F_{e}$ transition.

Because a $10$ G field is weak enough not to cause substantial hyperfine
level mixing (i.e., the magnetic sublevel splitting in the magnetic field
still is small in comparison with hyperfine splitting,) the increase in
orientation for $F_{i}=2$ absorption can be understood using the relative
transition probability $W_{F_{i}\longrightarrow F_{e}}$ given by Sobelman 
\cite{sob1}: 
\begin{equation}
W_{F_{i}\longrightarrow
F_{e}}=(2F_{i}+1)(2F_{e}+1)(2J_{i}+1)(2J_{e}+1)\left\{ 
\begin{array}{ccc}
J_{i} & F_{i} & I \\ 
F_{e} & J_{e} & 1
\end{array}
\right\} \left\{ 
\begin{array}{ccc}
L_{i} & J_{i} & S \\ 
J_{e} & L_{e} & 1
\end{array}
\right\} ^{2}.  \label{eq5}
\end{equation}
Here $J_{i},J_{e}$ and $L_{i},L_{e}$ are quantum numbers of total and
orbital electronic angular momentum of the initial and final atomic state
and $S$ is the electronic spin of the atomic state. Quantities in curled
brackets are $6-j$ symbols. \ This expression predicts that the $%
F_{i}=2\longrightarrow F_{e}=2$ absorption contributes $39\%$ of the total
allowed ( $\Delta F=0,\pm 1$) absorption from $F_{i}=2.$ \ In contrast, the $%
F_{i}=3\longrightarrow F_{e}=2$ absorption contributes only $8\%$ of the
total allowed ($\Delta F=0,\pm 1$) absorption from $F_{i}=3$. At the same
time the $F_{e}=2$ state is the state for which the magnetic sublevels
undergo a level crossing in the vicinity of a $10$ G magnetic field. \ Thus
the absorption from the $F_{i}=2$ state leads to a greater degree of
transverse orientation.

Similar level splitting diagrams (Figure \ref{Rb_87_levels}) and expected
circularity signals (Figure \ref{Rb_87_circ}) are calculated also for
rubidium isotope $^{87}$Rb. In this case the following atomic constants are
used: $I=3/2$, $a=84.845$ MHz, $b=12.52$ MHz, $g_{J}=-1.3362$, $%
g_{I}=-0.000995$ \cite{ari1}.

\section{Experimental}

In our experiment we use isotopically enriched rubidium (99 \% of $^{85}$Rb)
contained in a glass cell at room temperature to keep atomic vapor
concentration low and avoid reabsorption. The $5s$ $^{2}S_{1/2}$ to $5p$ $%
^{2}P_{3/2}$ transition at $780.2$ nm is excited using a temperature- and
current- stabilized single-mode diode laser (Sony SLD114VS). \ Absorption
signal is measured using a photodiode. \ As the laser frequency is swept
using a ramped current drive, two absorption peaks with half-width of about $%
600$ MHz separated by $\sim 3$ GHz appear due to the $^{85}$Rb ground state
hyperfine structure. The excited-state hyperfine structure is not resolved
under the Doppler profile and introduced laser-frequency jittering. \ The
laser line width without jittering is about $60$ MHz. To avoid optical
pumping and other nonlinear effects, neutral density filters are used to
reduce the laser intensity until absorption lines at $60$ G broaden by less
than 10\%.

During the level crossing and circularity measurements, the laser wavelength
is stabilized on one of the two absorption peaks. Fluorescence is monitored
on an axis normal to the electric vector ${\bf E}_{exc}$ and external
magnetic field ${\bf B}$. A two-lens system is used to image the
fluorescence on a photodetector containing a $3\times 3$ mm photodiode
(Hamamatsu S1223-01) and a transimpedance amplifier. A rotating ($f=240$ Hz)
sheet polarizer is inserted between the lenses. The photodetector signal is
fed to a lock-in amplifier (Femto LIA-MV-150) that measures the intensity
difference of two orthogonal linearly polarized fluorescence components. A
magnetic field of up to $65$ G is produced by passing current through a pair
of Helmholz coils $20$ cm in diameter. The uncertainty of the magnetic field
is estimated to $\pm 0.3$ G. \ The sweep time is $5$ s and $256$ sweeps are
averaged on an IBM\ compatible computer with a National Instruments data
acquisition card. A lock-in time constant of $10$ ms is used. Several
adjustments are made to record symmetrical level crossing signals while
sweeping the magnetic field in opposite directions. First, a linear
polarizer is placed in a laser beam before the rubidium cell to fine adjust
the laser polarization. \ Second, the lock-in phase is adjusted and, third,
the Earth magnetic field components are compensated with additional Helmholz
coils.

To detect circularly polarized light the gain electronics are first adjusted
so that the linear polarization signals are symmetrical in opposite magnetic
field directions. \ A $\lambda /4$ wave plate is then placed before the
polarizer so that right- and left- handed circularly polarized light
components are converted to opposite polarizations. \ It is checked that
circularity signal at $B=0$ is zero. \ During the circularity measurements
the magnetic field is swept alternatively in one and another direction and
both traces are averaged. \ The experimentally recorded signal actually is $%
I({\bf E}_{right})-I({\bf E}_{left})$ and not the ratio $(I({\bf E}%
_{right})-I({\bf E}_{left}))/(I({\bf E}_{right})+I({\bf E}_{left}))$. \
Numerical simulations reveal that these two signals have almost the same
shapes, the relative difference is less than $3\%$. Experimentally recorded
signals are scaled vertically to fit the calculated ones.

Figures \ref{Rb_85_circ} and \ref{Rb_87_circ}\ compare experimentally
obtained circularity to the theoretical ones. After the scaling factor to
the experimental signal is applied (no other adjustable parameters are used)
an excellent agreement between theoretical predictions and experimental
signals can be observed. For both isotopes circularity signals with
amplitude of several percents are measured.

In case of the measurements with $^{87}$Rb another cell was used, that
contained isotopically enriched $^{87}$Rb ($99$\%). For this isotope the
signal starting from $F_{i}=1$ exhibits stronger resonance circularity than
the one starting from the $F_{i}=2$ ground state. The reason for this is the
same as already discussed in a Section \ref{se_th} for the $^{85}$Rb
isotope. Only in this case the excited state hyperfine component $F_{e}=1$
undergoes level crossings with $\Delta m_{F_{e}}=\pm 1$. This resonance
intensity ratio for two measured signals reflects the general situation that
transitions with $\Delta F=0$ are more intense than transitions with $\Delta
F=\pm 1$.

\section{Comparison of circularity measurements to other level-crossing
measurements}

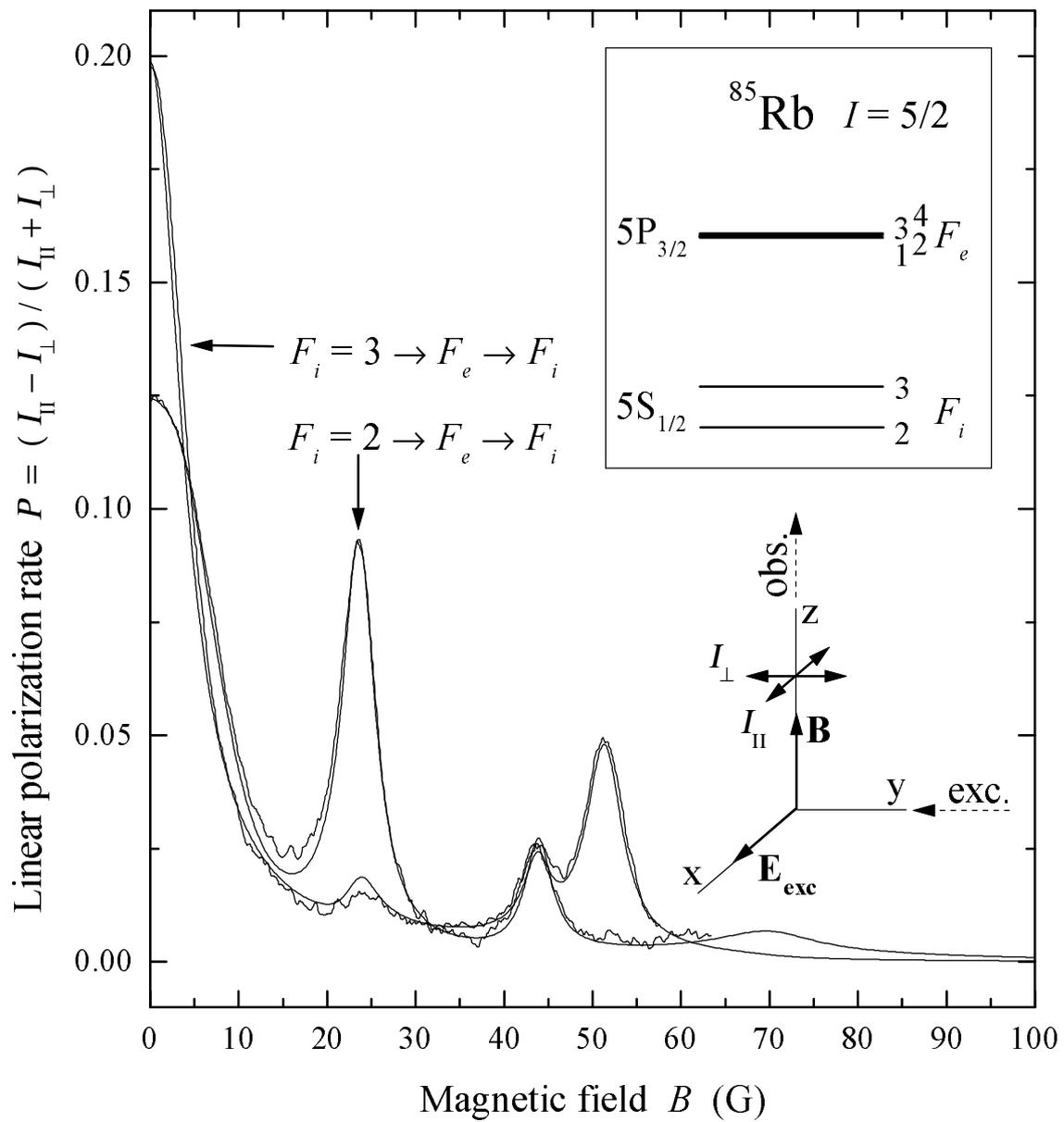
\begin{figure}[tbp]
\caption{Numerically simulated (smooth line) and experimentally measured
(signal with noise) level crossing signals in linearly polarized
fluorescence for $^{85}$Rb.}
\label{Rb_85_lin}
\end{figure}

In previous studies, atoms are excited by a linearly polarized light with%
{\bf \ }${\bf E}_{exc}${\bf \ }vector perpendicular to an external magnetic
field. The fluorescence emitted along the magnetic field is then detected.
Fluorescence linear polarization as a function of magnetic field is
measured. \ Here we repeat this experiment for the case of $^{85}$Rb (see
inset of Figure \ref{Rb_85_lin}.) Two signals are numerically simulated and
experimentally recorded, the first one when absorption occurs on the
transitions ($F_{i}=2\rightarrow F_{e}$) and a second one for a ($%
F_{i}=3\rightarrow F_{e}$) absorption transition. In both cases the
conditions are maintained so that the excited state hyperfine levels are not
resolved.\ \ For the first absorption transition in the absence of the
magnetic field electric dipole transitions are allowed only to the levels $%
F_{e}=1,2,$ and $3$. For the second absorption transition in absence of the
external field hyperfine components with $F_{e}=2,3,$ and $4$ can be excited.

In presence of the magnetic field selection rules change substantially. As
it was mentioned before, $F_{e}$ is no longer a good quantum number. Each
hyperfine level in the presence of external field is mixed together with
others. As far as $m_{F_{e}}$ remains a good quantum number in the presence
of the external field, only components with the same $m_{F_{e}}$ are mixed.\
\ This implies that for the present example of $^{85}$Rb, magnetic sublevels
with $m_{F_{e}}=4$ and $-4$ at any field value are unmixed because only $%
F_{e}=4$ contains such sublevels and there is no counterpart for these
states to be mixed with. In case of $m_{F_{e}}=3$ and $-3$ only two magnetic
sublevels originating from $F_{e}=3$ and $4$ are mixed together, etc. This
means that magnetic sublevels $m_{F_{e}}=0,\pm 1$ in external field are
composed from $F_{e}=1,2,3,4$ components, $m_{F_{e}}=\pm 2$, from $%
F_{e}=2,3,4$ components $m_{F_{e}}=\pm 3$, from $F_{e}=3,4$ components but $%
m_{F_{e}}=\pm 4$, contain only one component $F_{e}=4$.

In Figure \ref{Rb_85_levels} we can see several $\Delta m_{F_{e}}=2$
magnetic sublevel crossings. \ The first crossing takes place at zero
magnetic field when all magnetic sublevels belonging to the same hyperfine
level have the same energy. \ This crossing is the zero field level
crossing. \ Because all magnetic sublevels belonging to the same hyperfine
state cross at zero field, the zero field level crossing leads to the
largest resonance amplitude. Then subsequent crossings take place at
approximately $2.4,4.2,8.2,24,44,52$ and $74$ G magnetic field strength. \ A
resonant peak occurs on both linear polarization signals for almost every
one of these level crossings, although with differing amplitudes (see Figure 
\ref{Rb_85_lin}.) There is one exception. The strong resonance peak at $52$
G that is present in $F_{i}=3\rightarrow F_{e}$ signal is missing in $%
F_{i}=2\rightarrow F_{e}$ signal. This seeming inconsistency can be easily
explained. This resonance appears when the magnetic sublevels $%
m_{F_{e}=4}=-4 $ and $m_{F_{e}=3}=-2$ are crossing. But as it was mentioned
due to dipole transition selection rules the $m_{F_{e}=4}=-4$ level can not
be excited from $F_{i}=2$ and this restriction can not be removed by
external field because $m_{F_{e}=4}=-4$ remains unmixed at any field
strength.

In the same Figure \ref{Rb_85_lin} along with the theoretically simulated
signal the experimentally registered signal is depicted as well. The only
adjustable parameter in this comparison is a scaling factor for the overall
intensity of the experimentally detected signal. The observed signals agrees
very well with level crossing signal registered by several groups before us 
\cite{sva1,sch1}. However, in these previous studies the first derivative
from the intensity was measured and so we were able to compare only the
exact positions of resonances. These coincide perfectly. In our case we are
able to calculate not only the positions of resonances, but also the shape,
width and relative amplitudes of the resonance peaks.

\section{Conclusions}

In this study we report for the first time the appearance of transverse
orientation in atoms with hyperfine structure after excitation by linearly
polarized light in the presence of an external magnetic field. \ We have
also presented a theory that is in\ quantitative agreement with our data.
The use of transverse alignment-orientation as a probe of level crossings is
compared to previous measurements \ The case of measured circularity has an
advantage over the more conventional measurement of the degree of linear
polarization:\ \ For the case of alignment -- orientation conversion, there
is no signal in the absence of the external field. \ This implies that we do
not have the first trivial resonance position at zero field value which is
always present in traditional geometry (Hanle effect \cite{mor1}.) This
allows measurements of first level crossing positions that are very close to
the zero field resonance. In traditional methods these resonances are hidden
under the zero field peak. For example, from the inset of Figure \ref
{Rb_85_levels} we can see that there must exist several resonances of $%
\Delta m_{F_{e}}=2$ crossings around $3$ and $6$ G. However these resonances
are hidden in a traditional level crossing signal and can not be observed
(see Figure \ref{Rb_85_lin}.) At the same time $\Delta m_{F_{e}}=1$
crossings that appear even at smaller field values $2,4$ and $8$ G in
alignment -- orientation conversion signals, although not fully resolved are
clearly visible. The possibility to detect these resonances can improve the
precision of atomic hyperfine splitting constants.

\section{Acknowledgments}

This work could not be completed without valuable comments and advice of Dr.
Habil. Phys. Maris Tamanis from Institute of Atomic Physics and
Spectroscopy, University of Latvia and continuous support of Prof. Sune
Svanberg from Department of Physics at Lund Institute of Technology. We are
thankful to Prof. Neil Shafer-Ray from University of Oklahoma \ for careful
reading of the manuscript and valuable comments. Financial support from
Swedish Institute Visby program is greatly acknowledged.

\begin{figure}[tbp]
\caption{Hyperfine structure energy-level diagram of $^{85}$Rb $%
5p^{2}P_{3/2} $ in an external magnetic field. Symbols $\square $ indicate $%
\Delta m_{F_{e}}=1$ level crossings, $\bigcirc $ --- $\Delta m_{F_{e}}=2$
level crossings.}
\label{Rb_85_levels}
\end{figure}

\begin{figure}[tbp]
\caption{Numerically simulated (smooth line) and experimentally measured
(signal with noise) level crossing signals in fluorescence circularity for $%
^{85}$Rb in conditions of alignment -- orientation conversion and production
of transversal orientation.}
\label{Rb_85_circ}
\end{figure}

\begin{figure}[tbp]
\caption{Hyperfine structure energy-level diagram of $^{87}$Rb $%
5p^{2}P_{3/2} $ state in an external magnetic field. Symbols $\square $
indicate $\Delta m_{F_{e}}=1$ level crossings, and $\bigcirc $ --- $\Delta
m_{F_{e}}=2$ level crossings.}
\label{Rb_87_levels}
\end{figure}

\begin{figure}[tbp]
\caption{Numerically simulated (smooth line) and experimentally measured
(signal with noise) level crossing signals in fluorescence circularity for $%
^{87}$Rb in conditions of alignment -- orientation conversion and production
of transversal orientation.}
\label{Rb_87_circ}
\end{figure}

\begin{references}
\bibitem{alignment}  In this work the terms alignment, orientation, and
polarization take on their technical meaning. \ Alignment refers to a
distribution of angular momentum direction with respect to an axis, but not
a preferred direction. \ Orientation refers to a distribution of angular
momentum direction that favors a given direction.
\end{references}
\end{document}